\documentclass[conference,a4paper]{APSIPA2025}
\usepackage{amsmath}
\usepackage{graphicx}
\usepackage{multirow}
\usepackage{threeparttable}
\usepackage[backend=biber,style=ieee,doi=false,isbn=false,url=false]{biblatex}
\usepackage{pgfplots}
\pgfplotsset{compat=newest}
\usepackage{multirow}
\usepackage{hhline}
\usepackage[normalem]{ulem}
\usepackage{balance}

\usepackage{geometry}
\usepackage{fancyhdr}

\fancypagestyle{firststyle}{
  \fancyhf{}
  \fancyhead[C]{2025 Asia Pacific Signal and Information Processing Association Annual Summit and Conference (APSIPA ASC)}
}

\begin{document}

\title{Leveraging Language Information for \\ Target Language Extraction}

\author{
\authorblockN{
Mehmet Sinan Yıldırım\authorrefmark{1},
Ruijie Tao\authorrefmark{1},
Wupeng Wang\authorrefmark{1},
Junyi Ao\authorrefmark{2},
Haizhou Li\authorrefmark{1}\authorrefmark{2}
}

\authorblockA{
\authorrefmark{1}
Department of Electrical and Computer Engineering, National University of Singapore, Singapore \\
E-mail: sinan@u.nus.edu}
\authorblockA{
\authorrefmark{2}
 School of Artificial Intelligence, Shenzhen Research Institute of Big Data, \\ The Chinese University of Hong Kong, Shenzhen, China}

}

\maketitle
\thispagestyle{firststyle}

\begin{abstract}
  Target Language Extraction aims to extract speech in a specific language from a mixture waveform that contains multiple speakers speaking different languages. The human auditory system is adept at performing this task with the knowledge of the particular language. However, the performance of the conventional extraction systems is limited by the lack of this prior knowledge. Speech pre-trained models, which capture rich linguistic and phonetic representations from large-scale in-the-wild corpora, can provide this missing language knowledge to these systems. In this work, we propose a novel end-to-end framework to leverage language knowledge from speech pre-trained models. This knowledge is used to guide the extraction model to better capture the target language characteristics, thereby improving extraction quality. To demonstrate the effectiveness of our proposed approach, we construct the first publicly available multilingual dataset for Target Language Extraction. Experimental results show that our method achieves improvements of 1.22 dB and 1.12 dB in SI-SNR for English and German extraction, respectively, from mixtures containing both languages.
\end{abstract}

\section{Introduction}
Humans have the extraordinary ability to focus on speech from a particular auditory source in a noisy environment with various interfering sources. This ability is known as the cocktail-party effect \cite{cherry1953some}. Inspired by this phenomenon, enabling machines to replicate this selective auditory perception has long been a significant research challenge. Recent advances in deep learning have led to a paradigm shift, enabling the development of high-quality models for this task.
\begingroup
\renewcommand\thefootnote{}%
\footnote{This research is supported by: Shenzhen Science and Technology Program (Shenzhen Key Laboratory, Grant No. ZDSYS20230626091302006); Shenzhen Science and Technology Research Fund (Fundamental Research Key Project, Grant No. JCYJ20220818103001002); Program for Guangdong Introducing Innovative and Entrepreneurial Teams, Grant No. 2023ZT10X044.}%
\addtocounter{footnote}{-1}%
\endgroup

Target speech extraction models are a common solution designed to mimic this selective auditory perception capability. These models are generally conditioned on an additional input from which they obtain the speaker characteristics. This auxiliary input is usually a reference utterance produced by the target speaker \cite{zmolikova2019speakerbeam, wang19hvoicefilter, zeng2023sefnet, sato24speakerbeamss}, but it can alternatively take the form of textual cues \cite{huoSpeakerIdentityText2025, liu2025separatetext, kim2025contextual, huo2025textguided}, visual cues \cite{tao2025avtse, zhaoxi2024avtselip, chung20facefilter}, or even EEG signals \cite{desilva2024NeuroSpex}.

\begin{figure}[t]
\begin{center}
\includegraphics[width=\columnwidth]{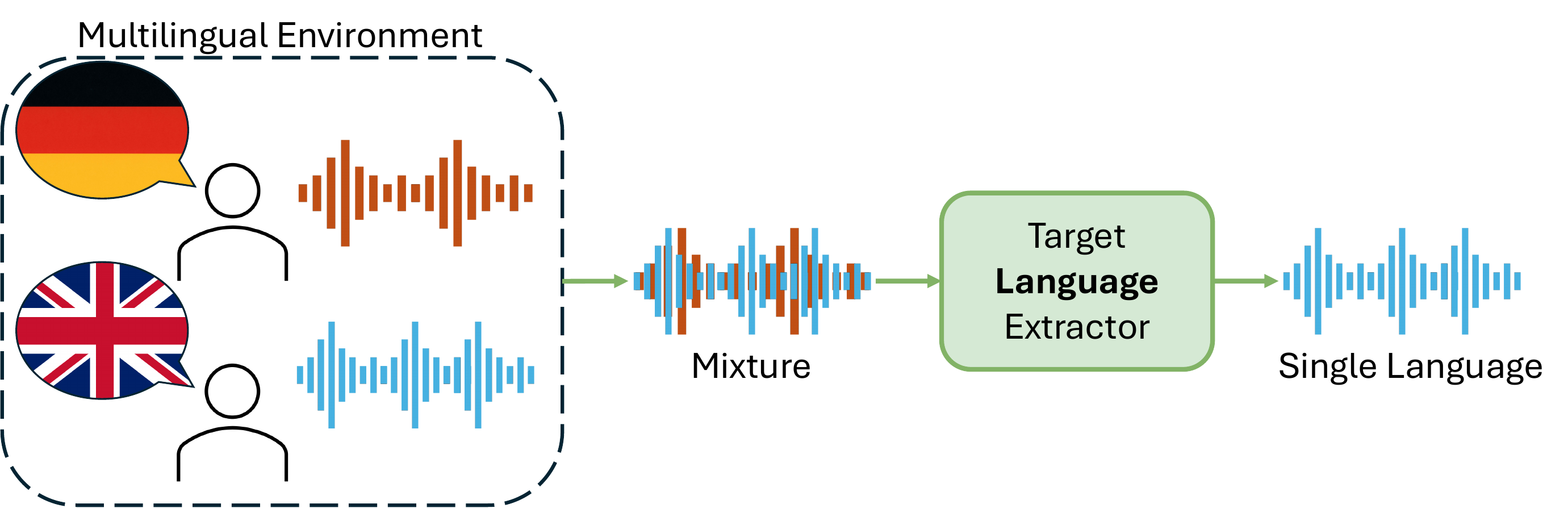}
\end{center}
\caption{Overview of the Target Language Extraction (TLE) task: given a speech mixture containing multiple languages (e.g. English and German), the model extracts only the speech corresponding to the target language (e.g., English), regardless of speaker identity.}
\label{fig:cover}
\end{figure}

The aforementioned cues are typically used in monolingual settings. However, real-world environments often involve multiple languages. For example, airport announcements are often delivered in several languages, and international participants at academic conferences frequently interact in different languages. In such scenarios, the user may be interested in extracting speech from a specific language. To address this need, the Target Language Extraction (TLE) task (as shown in Fig.~\ref{fig:cover}) was introduced~\cite{borsdorfTargetLanguageExtraction2021} with the aim of isolating speech corresponding to a particular language from multilingual conversations. This task can serve as an essential preprocessing step in multilingual settings, enabling downstream systems to process each language separately.

To achieve TLE, prior works \cite{borsdorfTargetLanguageExtraction2021, borsdorfExpertsAllRoundersTarget2022, borsdorfBlindLanguageSeparation2022} primarily follow approaches developed for Target Speaker Extraction (TSE), adapting these models to extract languages rather than speakers. Representative models include SpEx+ \cite{ge20_spexplus}, Conv-TasNet \cite{luo2019convtasnet}, and SepFormer \cite{subakan2021attention}. However, these methods largely ignore the inherent linguistic structure of the target language. Studies in auditory perception suggest that humans recognize and extract speech more effectively in familiar languages than in unfamiliar ones \cite{bradlow1999nativerecognition}, due to prior exposure to the language's phonological and syntactic patterns. Natural languages exhibit unique phonetic and linguistic characteristics. Familiarity with these characteristics can aid with extraction and help mitigate potential errors during the extraction process.

Motivated by the role of linguistic knowledge in human auditory perception, we hypothesize that injecting language information into the extraction network can enhance the overall quality of TLE. A key challenge is how to incorporate such language information effectively into the extraction process. Recent advances in self-supervised speech representation learning provide a promising solution, as these models are trained on large-scale unlabeled data and are capable of capturing universal, robust speech and language representations~\cite{hsu2021hubert, chen2022wavlm, baevski2020wav2vec}. Large-scale pre-trained speech models have demonstrated a strong capacity for language modeling, capturing not only phonetic features but also higher-level linguistic structures. To facilitate TLE with language information, we introduce a novel approach that employs an auxiliary training loss. This loss aligns the internal representations of the extraction model with those from a pre-trained speech model. No additional computation is incurred during inference since this auxiliary supervision is only applied during training.

In addition, previous TLE studies \cite{borsdorfTargetLanguageExtraction2021, borsdorfExpertsAllRoundersTarget2022, borsdorfBlindLanguageSeparation2022} constructed their mixture datasets using samples from the GlobalPhone corpus \cite{schultz2013globalphone}, which is not publicly available. To enable a fair comparison on an open-source benchmark, we simulate a new bilingual dataset following the same principles using the CommonVoice~\cite{ardilaCommonVoiceMassivelyMultilingual2020} dataset which is publicly available.


In summary, our contributions are three-fold:
\begin{itemize}
    \item We propose a framework that injects language information into the TLE pipeline by leveraging pre-trained speech models.
    \item We provide the first publicly available dataset for TLE, named CommonVoiceMix
    \item We demonstrate that our framework enhances the extraction performance for TLE and establish it as the first publicly available baseline for future studies
\end{itemize}

The remainder of this manuscript is organized as follows. In section II, we present a formal definition of TLE, the design of the combined loss function, the newly constructed dataset, and the experimental setup. Section III presents the experimental results and their interpretations. Section IV concludes this work.

\begin{figure}[t]
\begin{center}
\includegraphics[width=\linewidth]{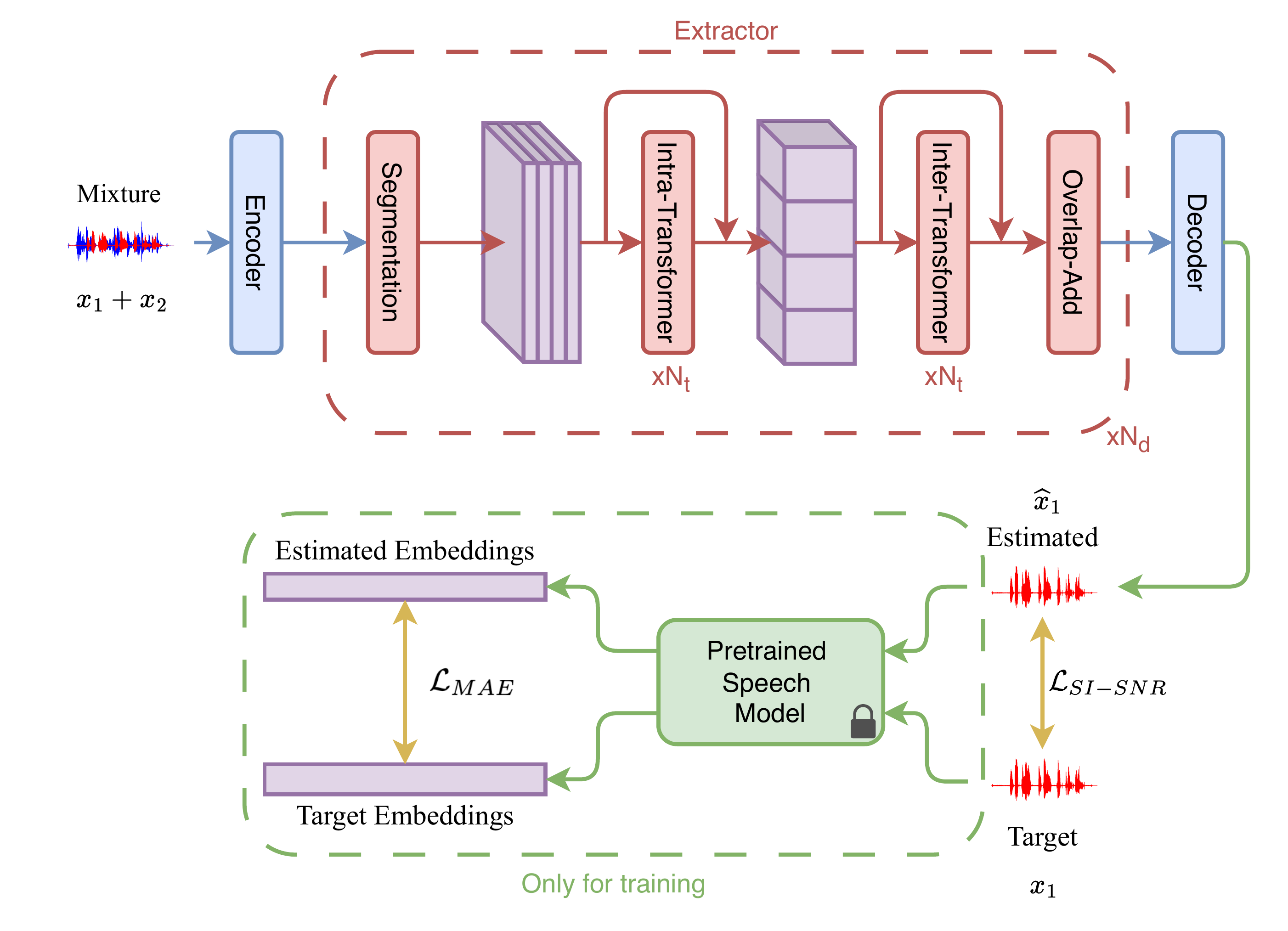}
\end{center}
\caption{Overview of the proposed method. Pre-trained speech model obtains embeddings for both the estimated and the target waveforms. An additional similarity loss between them contributes to align the internal speech representations of the base model}
\label{fig:baseline}
\end{figure}

\ 

\section{Target Language Extraction with \\ Language Information}
\subsection{Target Language Extraction}

Target Language Extraction is the task of isolating speech in a specific target language from a mixture containing speech in multiple languages by different speakers. Let $x_i(t)$ denote the speech waveform for language $i$. Then the mixture signal $m(t)$ containing speech from $N$ languages is given by:
$$m(t) = \sum_{i=1}^{N} x_i(t)$$

TLE aims to recover the original target language speech from this mixture. This can be formulated as finding an optimal transformation $f(\cdot)$ such that the estimation
$$\widehat{x_i}(t) = f(m(t))$$
is as close to the original target language speech $x_i(t)$ as possible according to a certain distance metric.


Previous works proposed using the scale-invariant signal-to-noise ratio (SI-SNR) \cite{le2019sdr}. We adopt this as our baseline loss. If the target language is language $i$, the SI-SNR loss is defined as
$$\mathcal{L}_{SI-SNR} = -10\log_{10}\left( \frac{||\alpha x_i(t)||^2}{||\alpha x_i(t) - \widehat{x_i}(t)||^2}\right) $$
where
$$\alpha = \frac{\widehat{x_i}(t)^T x_i(t)}{||x_i(t)||^2}.$$

The SI-SNR loss ensures that the estimated output closely matches the target output in terms of signal fidelity. 

\subsection{Proposed Method for Injecting Language Information}
This vanilla TLE approach does not explicitly use any intrinsic information of the underlying natural language, as it only optimizes for fidelity, although leveraging such information could benefit the extraction process. Self-supervised speech models capture this information as they are trained on large speech corpora to learn latent representations for speech signals. Since we work with multiple languages, it is more beneficial to use models trained on multilingual corpora that include the target languages in our dataset.

To achieve this goal, we introduce a TLE model that incorporates language knowledge as supervision to better guide the extraction process. Our model is inspired by the Sepformer \cite{subakan2021attention}, which has shown strong performance on monaural speech processing. As shown in Fig.~\ref{fig:baseline}, the mixture speech containing multiple languages is processed with a 1-D convolutional encoder. Then the encoder output is segmented along the time dimension into chunks of fixed length. An Intra-Transformer module applies attention within each chunk to model local dependencies. Subsequently, an Inter-Transformer module applies attention across corresponding time points of different chunks to capture global dependencies. The segments are then recombined and decoded to obtain the target speech estimate $\widehat{x_i}(t)$.

We process both the target signal $x_i(t)$ and the estimated signal $\widehat{x_i}(t)$ through a self-supervised pre-trained speech model to obtain their latent representations. These representations are rich in contextual information learned from the dependencies in the pre-training corpus. Ideally, these two representations should be as similar as possible. To enforce this similarity, we define a distance function between them as an auxiliary loss to guide the model toward extracting speech with correct linguistic structure and thus achieving better language understanding. We use the L1 distance to this end. Since the SI-SNR loss is in logarithmic scale, we also take the logarithm of the L1 distance to match the scale and sensitivity of the losses. Therefore, given a certain pre-trained model $h(\cdot)$, we have:
$$\mathcal{L}_{MAE} = 10\ \log_{10}\left(\frac{1}{d} \sum_{j=1}^d \bigg|[h(x_i(t))]_j - [h(\widehat{x_i}(t))]_j\bigg|\right)$$
where $d$ is the dimension of the pre-trained model output. We add this loss to the original SI-SNR loss with a coefficient $\beta \geq 0$.
$$\mathcal{L} = \mathcal{L}_{SI-SNR} + \beta \mathcal{L}_{MAE}$$

$\beta = 0$ corresponds to the baseline case of only using SI-SNR loss. The overall architecture is illustrated in Fig.~\ref{fig:baseline}. We train the model in two stages: first with $\beta = 0$ to learn general extraction, and then with a positive $\beta$ value to refine the extraction using language information. Note that this auxiliary supervision does not incur additional computational cost at inference time since it is only used during training.



\subsection{CommonVoiceMix: A New Public Corpus for TLE} 
To evaluate our method, we require a suitable multilingual dataset. In the TSE literature, commonly used datasets such as LibriMix \cite{cosentino2020librimix} and wsj0-2mix \cite{hershey2016wsj0-2mix} generate mixture signals from the LibriSpeech \cite{2015panayotovLibrispeech} and WSJ0 \cite{paul1992wsj} corpora, respectively. We adopt the LibriMix strategy to construct our dataset.

To construct multilingual mixtures, we need a base dataset that contains utterances in different languages and is ideally publicly available. Additionally, it should contain a high number of speakers to ensure that the method focuses on language rather than speaker identity. Due to these reasons, we chose the publicly available CommonVoice dataset \cite{ardilaCommonVoiceMassivelyMultilingual2020}. The appropriate mixtures are obtained by mixing samples from different languages. We chose English and German for our experiments. We discarded samples shorter than 7 seconds. The speaker IDs are utilized to verify that there is no speaker overlap between the train, dev, and test sets of the two languages.


With this strategy, we constructed the CommonVoiceMix dataset that comprises 30000, 4600 and 4500 English-German mixtures for train, dev, and test sets, respectively. We limit the number of training samples to 30000 whereas dev and test sets are limited by the number of samples of the lower-resource language in the original dataset. We randomly sample 6-second segments from the train mixtures such that the target speech is not empty. The entire utterance is used for dev and test sets. This corresponds to approximately 50 hours, 11.2 hours, and 11.2 hours for train, dev, and test sets, respectively. We also utilize the client\_id values of the speakers to identify the number of unique speakers in the dataset. A summary can be seen in Table~\ref{tab:dataset}.

We believe that a public TLE dataset will promote and accelerate research in this field. By using the same dataset, new approaches can directly compare to our work, which can serve as a baseline. For this purpose, our code and metadata are publicly available. \footnote{https://github.com/msinanyildirim/CommonVoiceMix} Other researchers can use these to construct our dataset and also create mixtures of other language combinations available in the CommonVoice dataset.



\begin{table}
\begin{center}
\caption{A Summary of the CommonVoiceMix dataset}
\label{tab:dataset}
\begin{tabular}{cccc}
\hline 
Subset & train & dev & test \\ \hline \hline
Total length (h) & 50 & 11.2 & 11.2 \\ \hline 
\# of samples & 30000 & 4600 & 4500 \\ \hline 
\# of unique English speakers & 9945 & 3126 & 4000 \\ \hline 
\# of unique German speakers & 4203 & 2276 & 2739 \\ \hline 
\end{tabular}
\end{center}
\end{table}

\subsection{Experimental Setup}

\cite{borsdorfTargetLanguageExtraction2021} shows that the best baseline for TLE is a Sepformer-based model \cite{subakan2021attention}. We use a dual-path transformer-based Sepformer model implemented in Speechbrain \cite{speechbrain} using a single mask to generate a single output. We use the default parameter values defined in Speechbrain \footnote{https://github.com/speechbrain/speechbrain}. However, we set the number of dual computing blocks to 1 due to computational constraints. This corresponds to $N_t=8$ and $N_d=1$ in Fig.~\ref{fig:baseline}. Additionally, the batch size is 2 during training in our experiments. This model serves as the extraction pipeline in our work.

For the self-supervised pre-trained model, there are several options such as HuBERT \cite{hsu2021hubert}, WavLM \cite{chen2022wavlm}, and wav2vec 2.0 \cite{baevski2020wav2vec}. As we deal with multiple languages, we choose mHuBERT-147 \cite{ zanonboitoMHuBERT147CompactMultilingual2024}, which is a HuBERT-based self-supervised learning model trained on 147 languages, including both English and German. mHuBERT-147 has the same size as the base HuBERT model. We use the output of the last layer, i.e. the 12th layer.

We use an Adam optimizer with an initial learning rate of $3\cdot10^{-4}$. We reduce the learning rate by half if the loss has not decreased for three consecutive epochs. We stop the training if the loss has not decreased for 20 epochs in a row. Initially, we start with $\beta=0$. This usually converges in around 80 epochs, which serves as the baseline. Subsequently, we reinitialize the optimizer, set $\beta=1$, and restart training. This second stage takes around 40 epochs to converge. 

\section{Experiments and Results}
\subsection{Main results}
The main experiments are performed as described, with English and German as the target language, respectively. We evaluate the performance in terms of SI-SNR, PESQ \cite{rix2001pesq} and STOI \cite{taal2010stoi}, which respectively assess the fidelity, the perceptual quality, and the short-term intelligibility of the estimated signals. For all three metrics, a higher value means a better quality. Table \ref{tab:mainresults} summarizes the results for English and German. For both languages, the second stage of multitask learning with the mHuBERT-147 significantly improves the results. The SI-SNR results are increased by 1.22 dB and 1.12 dB for English and German extractions, respectively. Similarly, STOI increases by 0.2 and 0.14, while PESQ increases by 0.05 and 0.03, respectively.

\begin{table}
\begin{center}
\caption{Main results on the CommonVoiceMix dev and test sets with mHuBERT-147 as the speech model. $\uparrow$ means higher is better.}
\label{tab:mainresults}
\begin{tabular}{cccccc}
\hline 
 Set & Target Lang & Method & SI-SNR (dB) $\uparrow$ & STOI $\uparrow$ & PESQ $\uparrow$ \\ \hline \hline
\multirow{4}{*}{dev}& \multirow{2}{*}{English} & baseline & 10.85 & 0.84 & 1.87 \\ 
& & \textbf{ours} & \textbf{12.22} & \textbf{0.87} & \textbf{2.08} \\ \cline{2-6} 
& \multirow{2}{*}{German} & baseline & 10.85 & 0.87 & 1.92 \\ 
& & \textbf{ours} & \textbf{11.98}  & \textbf{0.89} & \textbf{2.07} \\ \hline
\multirow{4}{*}{test}& \multirow{2}{*}{English} & baseline & 9.96 & 0.82 & 1.85 \\ 
& & \textbf{ours} & \textbf{11.18 }& \textbf{0.84} & \textbf{2.05} \\ \cline{2-6} 
& \multirow{2}{*}{German} & baseline & 10.03 & 0.84 & 1.91 \\ 
& & \textbf{ours} &  \textbf{11.15} & \textbf{0.87} & \textbf{2.05 }\\ \hline
\end{tabular}
\end{center}
\end{table}

These results support our hypothesis. In the second stage, the model optimizes a combined loss that includes the additional language-aware term. This is achieved partly by decreasing the $\mathcal{L}_{MAE}$. This occurs when the estimated signals produce latent embeddings similar to those of the target. Because the SSL model embeds language information, the estimated waveforms must be similar to the target waveform in terms of linguistic content and overall contextual consistency. Consequently, the extraction performance improves, as reflected in the higher SI-SNR values and higher PESQ values indicating enhanced perceptual quality.

It is important to note that the SSL model is frozen and not trained together with the core model. Additionally, the core model operations do not depend on the SSL parameters, as the SSL model only processes the output of the core model to calculate the auxiliary loss. Therefore, the SSL model is not necessary during inference. Thus, this performance upgrade with the better language understanding does not affect the inference speed in practical scenarios.

In addition, the overall second stage training does not depend on the specific SSL model used. Different SSL models could be used instead. We compare two models in the next section.

\subsection{Ablation studies}
\subsubsection{Using HuBERT}
Instead of mHuBERT-147, we can use the original HuBERT model, which was trained on a corpus containing English speech only. We keep all other parameter values the same. The results are presented in Table \ref{tab:hubertresults}. The SI-SNR improvement over the baseline is comparable to that achieved with the mHuBERT-147 for English. However, for German, the performance gain with HuBERT is slightly less than that with the mHuBERT-147. Nevertheless, there is still a significant performance gain compared to the baseline. This suggests that HuBERT still has fairly good representations for German speech, which helps the extraction. An SSL model trained on speech from both languages is still better. This can be likened to two people: one who only speaks English and another who speaks multiple languages, including English and German. The person who only speaks English can still focus on German speech to some extent, but it is not as effective as the multilingual person.

\begin{table}
\begin{center}
\caption{Results on CommonVoiceMix test set with HuBERT as the pre-trained speech model. $\uparrow$ means higher is better.}
\label{tab:hubertresults}
\begin{tabular}{ccccc}
\hline 
 Target Lang & Method & SI-SNR (dB) $\uparrow$ & STOI $\uparrow$ & PESQ $\uparrow$ \\ \hline \hline
 \multirow{3}{*}{English} & baseline & 9.96 & 0.82 & 1.85 \\ 
  & \textbf{mHuBERT-147} & \textbf{11.18} & \textbf{0.84} & \textbf{2.05} \\ 
 & HuBERT & 11.09 & 0.84 & 1.97 \\ \hline 
 \multirow{3}{*}{German} & baseline & 10.03 & 0.84 & 1.91 \\ 
 & \textbf{mHuBERT-147} & \textbf{11.15} & \textbf{0.87} & \textbf{2.05} \\ 
 & HuBERT & 10.97 & 0.86 & 1.98 \\ \hline 
\end{tabular}
\end{center}
\end{table}

\subsubsection{Different $\beta$ values}
For our training strategy, $\beta$ is a hyperparameter that must be chosen carefully to achieve optimal performance. In our results presented in Table \ref{tab:betaresults}, we observe that $\beta\approx1$ gives a good performance overall. As $\beta$ deviates from this value, the performance degrades. However, this second stage training strategy still offers improvements over the baseline.



\begin{table}
\begin{center}
\caption{Results on CommonVoiceMix test set with mHuBERT-147 as the speech model for different $\beta$ values. $\uparrow$ means higher is better.}
\label{tab:betaresults}
\resizebox{\linewidth}{!}{
\begin{tabular}{c|ccc|ccc}
\hline 
\multirow{2}*{$\beta$} & \multicolumn{3}{c|}{English} & \multicolumn{3}{c}{German}\\ 
~ & SI-SNR (dB) $\uparrow$ & STOI $\uparrow$ & PESQ $\uparrow$ & SI-SNR (dB) $\uparrow$ & STOI $\uparrow$ & PESQ $\uparrow$\\ \hline \hline
  0.1 & 11.04 & 0.84 & 2.01 & 10.89 & 0.86 & 2.01 \\ 
 0.2 & 10.97 & 0.84 & 2.01 & 11.15 & 0.86 & 2.04 \\  
 \textbf{1} & \textbf{11.18} & \textbf{0.84} & \textbf{2.05} & \textbf{11.15} & \textbf{0.87} & \textbf{2.05} \\  
 5 & 10.83 & 0.84 & 2.00 & 10.83 & 0.87 & 2.03 \\  
 10 & 10.71 & 0.84 & 2.00 & 10.95 & 0.87 & 2.05 \\ \hline 

\end{tabular}
}
\end{center}
\end{table}

\subsubsection{Smaller model size} Since our strategy injects language information into the extraction process, it can help boost the extraction quality of smaller models. To test this, we try the same procedure for a smaller version of the model with $N_t=4$. This approximately halves the model size. The results are presented in Table \ref{tab:smallresults}. As expected, the smaller model exhibits lower performance due to its reduced parameter count. In this setting, our training strategy still benefits the extraction performance. However, the absolute SI-SNR gain is lower than in the larger model. This is attributed to the limited capacity of the smaller model, which restricts its ability to process rich representations.

\begin{table}
\begin{center}
\caption{Results on CommonVoiceMix test set using the model with half the size, $N_t=4$ instead of 8. $\uparrow$ means higher is better.}
\label{tab:smallresults}
\begin{tabular}{ccccc}
\hline 
 Target Lang & Method & SI-SNR (dB) $\uparrow$ & STOI $\uparrow$ & PESQ $\uparrow$ \\ \hline \hline
\multirow{2}{*}{English} & baseline & 8.17 & 0.79 & 1.68 \\ 
& \textbf{ours} & \textbf{9.10} & \textbf{0.81} & \textbf{1.83} \\ \cline{1-5} 
\multirow{2}{*}{German} & baseline & 8.76 & 0.83 & 1.74 \\ 
& \textbf{ours} & \textbf{9.66 }& \textbf{0.85} & \textbf{1.89} \\ \hline
\end{tabular}
\end{center}
\end{table}

\subsubsection{Different pre-trained model}
In our approach, different pre-trained models can be used in place of HuBERT, and each model may result in different extraction quality. As an example, we use WavLM \cite{chen2022wavlm} as the pre-trained speech model. A multilingual version of WavLM is not yet publicly available. Therefore, to ensure a fair comparison, we evaluate its performance against the monolingual HuBERT. Additionally, we use WavLM-base since it has a similar number of parameters to HuBERT. As presented in Table~\ref{tab:wavlmresults}, we observe that WavLM has a better performance than HuBERT. Our approach is adaptable to future models that may offer even better performance. Finally, a multilingual version of WavLM could potentially perform better, as suggested by the trends in Table \ref{tab:hubertresults}. However, such a model is not currently available for evaluation.

\begin{table}
\begin{center}
\caption{Results on CommonVoiceMix test set using WavLM as the pre-trained model. $\uparrow$ means higher is better.}
\label{tab:wavlmresults}
\begin{tabular}{ccccc}
\hline 
 Target Lang & Method & SI-SNR (dB) $\uparrow$ & STOI $\uparrow$ & PESQ $\uparrow$ \\ \hline \hline
\multirow{2}{*}{English} & HuBERT & 11.09 & 0.84 & 1.97 \\ 
& WavLM & 11.13 & 0.84 & 2.00 \\ \cline{1-5} 
\multirow{2}{*}{German} & HuBERT & 10.97 & 0.86 & 1.98 \\ 
& WavLM & 11.28 & 0.87 & 2.04 \\ \hline
\end{tabular}
\end{center}
\end{table}

\section{Conclusion}

In this work, we addressed the challenge of Target Language Extraction (TLE), which aims to isolate speech in a target language from multilingual mixtures. We introduced a novel training framework that injects language knowledge into the extraction pipeline by leveraging pre-trained self-supervised speech models. Specifically, we incorporated an auxiliary loss based on mean absolute error between latent representations derived from mHuBERT-147, guiding the model to better capture language-specific characteristics. We validated our approach on CommonVoiceMix, a new publicly available dataset we constructed for this task. Our experiments demonstrated that the proposed method consistently improves extraction quality, achieving notable SI-SNR and perceptual gains, while introducing no additional inference cost. Furthermore, we showed that pre-training on multilingual data enhances performance, highlighting the importance of language diversity in model design. In the future, we plan to investigate alternative pre-trained models, such as wav2vec 2.0, and extend this framework to causal, real-time scenarios. We believe that our approach provides a promising direction for advancing language-aware speech extraction systems.


\printbibliography

\end{document}